\documentclass[aps,prd,notitlepage,amssymb,amsmath,floatfix,nofootinbib,superscriptaddress]{revtex4-1}

\usepackage{epsfig}
\usepackage{bm}
\usepackage{amssymb}
\usepackage{amsmath}
\usepackage{color}
\usepackage{slashed}
\usepackage{subfigure}
\usepackage[colorlinks,linkcolor=blue,anchorcolor=black,citecolor=blue]{hyperref}
\allowdisplaybreaks[4]

\begin{document}

\title{Probing the longitudinal spin transfer via dihadron polarization correlations in unpolarized $e^+e^-$ and $pp$ collisions}

\author{Hao-Cheng Zhang}
\affiliation{Taishan College, Shandong University, Jinan, Shandong, 250100, China}

\author{Shu-Yi Wei}   
\email{shuyi@sdu.edu.cn}
\affiliation{Institute of Frontier and Interdisciplinary Science, Key Laboratory of Particle Physics and Particle Irradiation (MOE), Shandong University, Qingdao, Shandong 266237, China}

\begin{abstract}
The longitudinal spin transfer represents the probability density of producing longitudinally polarized hadrons from longitudinally polarized quarks or circularly polarized gluons. It thus was usually measured in polarized reactions or high-energy collisions where weak interaction dominates. In this work, we propose the dihadron polarization correlation as a novel probe of this quantity. Such an observable does not require the fragmenting partons to be polarized and therefore can be measured in the currently available experimental facilities, such as Belle, RHIC, Tevatron, and the LHC. We make quantitative predictions for these experiments. In light of the data already harvested, the experimental investigation of this observable provides more opportunity for the quantitative study of the longitudinal spin transfer. In particular, the measurements in $pp$ collisions can significantly constrain the fragmentation function of a circularly polarized gluon.
\end{abstract}

\maketitle

\section{Introduction}

Fragmentation functions (FFs) are essential non-perturbative inputs in making quantitative predictions for high-energy reactions \cite{Metz:2016swz}. While the unpolarized ones \cite{Binnewies:1994ju, deFlorian:1997zj, Kniehl:2000fe, Kneesch:2007ey, Kretzer:2000yf, Albino:2005me, Albino:2005mv, Albino:2008fy, deFlorian:2007aj, deFlorian:2007ekg, Hirai:2007cx, Aidala:2010bn, deFlorian:2014xna, deFlorian:2017lwf, Bertone:2017tyb, Khalek:2021gxf, arXiv:1905.03788, arXiv:2101.04664, arXiv:2210.06078} are well constrained for several hadrons by the experimental data from $e^+e^-$ annihilations, semi-inclusive deep inelastic collisions (SIDIS), and $pp$ collisions, the quantitative study on spin dependence is still in its early stage. The spin-dependent FFs are usually measured in polarized collisions or high-energy reactions where weak interaction dominates. It is very challenging to acquire experimental data with high accuracy. Recently, the spin-dependent FF was also investigated at Belle \cite{arXiv:1808.05000}, which is neither a polarized nor a weak-interaction-dominating experiment. This is possible since the Belle experiment measures the $D_{1T}^\perp$ FF, which represents the probability density of producing transversely polarized hadrons from unpolarized partons. As shown in Ref. \cite{arXiv:1808.05000}, the experimental data is accurate enough for phenomenological studies \cite{Anselmino:2019cqd, DAlesio:2020wjq, Callos:2020qtu, Kang:2020xyq, Li:2020oto, Chen:2021hdn, Gamberg:2021iat, DAlesio:2021dcx, DAlesio:2022brl}.

On the other hand, the longitudinal spin transfer $G_{1L} (z)$ represents the probability density of producing longitudinally polarized hadrons from longitudinally polarized partons. To measure this quantity in the experiment, one has to generate longitudinally polarized final state partons. Therefore, it is usually studied at LEP \cite{ALEPH:1996oew, OPAL:1997oem}, polarized SIDIS \cite{E665:1999fso, HERMES:1999buc, HERMES:2006lro, COMPASS:2009nhs}, NOMAD \cite{NOMAD:2000wdf, NOMAD:2001iup}, and RHIC \cite{STAR:2009hex, STAR:2018pps} experiments. LEP is an electron-positron collider at the $Z^0$-pole, which ensures that the fragmenting quarks are strongly polarized along the longitudinal direction \cite{Augustin:1978wf, Gustafson:1992iq}. The electron-positron annihilation process is a clean process to investigate FFs since it avoids the ``contamination'' from parton distribution functions (PDFs) \cite{Kotzinian:1997vd, deFlorian:1997zj, Liang:1997rt, Boros:1998kc,Ma:1998pd,Ma:1999gj,Ma:1999wp,Liu:2000fi,Chen:2016iey}. However, it receives little contribution from the gluon jet. This feature is double-edged. On one hand, it makes the quark signals cleaner at LEP. On the other, it provides little information on the gluon FF. Besides, HERMES \cite{HERMES:1999buc, HERMES:2006lro} and COMPASS \cite{COMPASS:2009nhs} also measured the longitudinal spin transfer coefficient in SIDIS with the polarized lepton beam, which also mainly probes the longitudinal spin transfer of quarks/antiquarks \cite{Ma:1999gj, Gluck:2000dy, Ma:2000ip, Ma:2000cg, Yang:2001yda, Leader:2001kh, Blumlein:2002qeu, Leader:2002az, Leader:2005ci}. Therefore, those experiments offer little constraint on the gluon longitudinal spin transfer ($G_{1L,g}$).  In principle, polarized $pp$ experiments can improve our understanding on this front \cite{STAR:2009hex, STAR:2018pps}. However, due to the large uncertainties in the experimental data and the ``contamination'' from the longitudinal spin transfer in PDFs ($g_{1L,g}$), the gluon spin transfer in FFs ($G_{1L,g}$) is still largely unknown. It is assumed to be negligible at the initial scale in the DSV parameterization  \cite{deFlorian:1997zj}. 

Moreover, the spin correlation has also been proposed as a probe to the longitudinal spin transfer \cite{Chen:1994ar} in unpolarized $e^+e^-$ collisions. The idea in that paper is intriguing. It is established on the fact that the helicities of the two partons produced in the same hard scattering are strongly correlated, although single-inclusive partons are {\it not} polarized. It thus proposed a genuine method to measure the longitudinal spin transfer with unpolarized final state partons. Such an idea deserves further investigation. In this work, we calculate the dihadron polarization correlation in unpolarized $e^+e^-$ annihilations and $pp$ collisions. We establish the connection between experimental observables and the longitudinal spin transfer. This work thus paves the way for a quantitative study of the longitudinal spin transfer at Belle, RHIC, Tevatron, and LHC experiments.

The paper is organized as follows. In Sec. II, we present our calculation for the dihadron polarization correlation in $e^+e^-$ annihilations. We use this simple process as an example to illustrate in detail the origin of the polarization correlation. Our approach is slightly different from that in Ref.~\cite{Chen:1994ar}, but in a more up-to-date language in QCD factorization. We also quantitatively predict observables at the Belle and LEP energies. In Sec. III, we show our corresponding results for the dihadron polarization correlation in unpolarized $pp$ collisions and make predictions for the observables at RHIC, LHC, and Tevatron energies. A summary is given in Sec. IV.

\section{Dihadron polarization correlation in $e^+e^-$ annihilations}

We consider the following process 
\begin{align}
e^+ + e^- \to \Lambda (\lambda_1, z_1) + \bar\Lambda (\lambda_1, z_2) + X,
\end{align}
with two almost back-to-back Lambda hyperons, where $\lambda_{1,2}$ denotes the helicity of the corresponding hadron, and $z_{1,2}$ represents the momentum fraction. The dihadron polarization correlation, ${\cal C}_{LL}$, is defined as the probability for $\lambda_1$ and $\lambda_2$ taking the same sign minus that taking opposite signs. The technical detail on how to measure it in the experiment is presented in the appendix. 

We use the photon exchange process as a simple example to illustrate our approach. The extension to the $Z^0$-boson exchange process and the $\gamma^* Z^0$-interference term is straightforward. At the leading order (LO) and leading twist (LT) approximation, the cross-section can be cast into \cite{Boer:1997mf, Wei:2014pma} 
\begin{align}
\frac{d\sigma^{\gamma\gamma}}{dy dz_1dz_2 d^2 \bm{P}_{\perp}}
= \frac{2\pi N_c \alpha^2}{Q^4} L_{\mu\nu} W^{\mu\nu},
\end{align}
where $N_c=3$ is the color factor, $y=(1+\cos\theta)/2$ with $\theta$ being the angle between incoming electron and outgoing $\Lambda$, $\bm{P}_{\perp}$ is the transverse momentum of $\bar \Lambda$ with respect to the $\Lambda$ momentum, and $Q$ is the center-of-mass energy of the colliding leptons. The leptonic tensor reads
\begin{align}
L_{\mu\nu} = l_{1\mu} l_{2\nu} + l_{1\nu} l_{2\mu} - g_{\mu\nu} l_1 \cdot l_2,
\end{align}
with $l_1$ and $l_2$ being the momenta of electron and positron respectively. The hadronic tensor can be written as \cite{Mulders:1995dh, Boer:1997mf, Boer:2003cm, Bacchetta:2006tn, Pitonyak:2013dsu, Wei:2014pma, Metz:2016swz} 
\begin{align}
W^{\mu\nu} = 
& \sum_q e_q^2 \int d^2 \bm{p}_{T1} d^2 \bm{p}_{T2} \delta^2 (\frac{z_2}{z_1} \bm{p}_{T1} + \bm{p}_{T2} - \bm{P}_{\perp})
\nonumber\\
& \times 
{\rm Tr} [
2 \hat \Xi_{q} (z_1, \bm{p}_{T1}) \gamma^\mu 2 \hat \Xi_{\bar q} (z_2, \bm{p}_{T2}) \gamma^\nu 
+
2 \hat \Xi_{q} (z_2, \bm{p}_{T2}) \gamma^\mu 2 \hat \Xi_{\bar q} (z_1, \bm{p}_{T1}) \gamma^\nu 
],
\end{align}
where $\Xi_{q/\bar q}$ is the quark-quark correlator that can be decomposed in terms of Gamma matrices. This is a standard procedure. A full decomposition at the LT level can be found in Refs. \cite{Mulders:1995dh, Boer:1997mf, Boer:2003cm, Bacchetta:2006tn, Pitonyak:2013dsu, Wei:2014pma, Metz:2016swz}. Here, we only keep two terms that are relevent to this study, namely the unpolarized FF and the longitudinal spin transfer. Therefore, we have
\begin{align}
&
\hat \Xi_{q} (z_1, \bm{p}_{T1}) = \frac{1}{4} \slashed n_+ D_{1, q} (z_1, \bm{p}_{T1}) + \frac{1}{4}  \gamma_5 \slashed n_+ \lambda_1 G_{1L, q} (z_1, \bm{p}_{T1}),
\\
&
\hat \Xi_{\bar q} (z_2, \bm{p}_{T2}) = \frac{1}{4} \slashed n_- D_{1, \bar q} (z_2, \bm{p}_{T2}) - \frac{1}{4}  \gamma_5 \slashed n_- \lambda_2 G_{1L, \bar q} (z_2, \bm{p}_{T2}),
\end{align}
where $n_\pm$ are unit vectors in the light-cone coordinate with plus direction being specified by the $\Lambda$ momentum, and $D_{1,q/\bar q} (z, \bm{p}_{T})$ and $G_{1L,q/\bar q} (z, \bm{p}_{T})$ are transverse momentum dependent FFs. There is a sign flip between $G_{1L, \bar q}$ and $G_{1L, q}$ terms. This is in line with that in Refs.~\cite{Mulders:1995dh, Boer:1997mf, Pitonyak:2013dsu} due to the charge conjugation symmetry. As shown bellow, this sign flip can be fully appreciated in the language of helicity amplitudes. It ensures that $G_{1L, \bar q}$ and $G_{1L, q}$ have the same physical interpretation. Following the same convention in Ref.~\cite{Boer:1997mf}, we have properly chosen the normalization of $G_{1L}$, so that $\lambda_{1,2} = \pm 1$ for spin-$1/2$ particles. Since the transverse momentum dependence is not the focus of this study, we integrate over $\bm{P}_{\perp}$ to simplify the discussion. Finally, we arrive at
\begin{align}
\frac{d\sigma^{\gamma\gamma}}{dydz_1dz_2} = \frac{2\pi N_c \alpha^2}{Q^2} \sum_q e_q^2 A(y) [D_{1,q} (z_{1})D_{1,\bar q} (z_{2}) - \lambda_1 \lambda_2 G_{1L,q}(z_{1}) G_{1L,\bar q}(z_{2})] +\{ q \leftrightarrow \bar q \},
\end{align}
where $A(y)= y^2 + (1-y)^2$ is the hard coefficient. Since we cannot distinguish a quark jet from an antiquark jet in the experiment, the exchange, $\{ q \leftrightarrow \bar q \}$, is implicit. Here, $D_{1,q/\bar q} (z) = \int d^2 \bm{p}_{T} D_{1, q/\bar q} (z, \bm{p}_{T})$ and $G_{1L,q/\bar q}(z) = \int d^2 \bm{p}_{T} G_{1L, q/\bar q} (z, \bm{p}_{T})$ are $p_T$-integrated FFs (or collinear FFs). This result agrees with that in Refs.~\cite{Boer:1997mf,DAlesio:2021dcx}.

After taking into account the $Z^0$-exchange contribution and the $\gamma^* Z^0$-interference term, we obtain the final cross section, which is given by
\begin{align}
\frac{d\sigma}{dy dz_1 dz_2} = \frac{2\pi N_c \alpha^2}{Q^2} \sum_q  & \Bigl[ \omega_q(y) \left[D_{1,q} (z_{1})D_{1,\bar q} (z_{2}) - \lambda_1 \lambda_2 G_{1L,q}(z_{1}) G_{1L,\bar q}(z_{2})\right]\nonumber\\
 +&\Delta \omega_q(y)[\lambda_1 G_{1L,q}(z_1) D_{1,\bar{q}}(z_2) - \lambda_2D_{1,q}(z_1)G_{1L,\bar{q}}(z_2)] \Bigr] + \{ q \leftrightarrow \bar q\}. \label{eq:cross-ee}
\end{align}
Here, the hard coefficients are given by $\omega_q(y)=\chi T_0^q(y)+e_q^2 A(y)+\chi_{\text {int }}^q I_0^q(y)$ and $\Delta\omega_q(y)=\chi T_1^q(y)+\chi_{\text {int }}^q I_1^q(y)$, with $T_0^q(y)=c_1^e c_1^q A(y)-c_3^e c_3^q B(y)$, $T_1^q(y)=-c_1^e c_3^q A(y)+c_3^e c_1^q B(y)$, $I_0^q(y)=c_V^e c_V^q A(y)-c_A^e c_A^q B(y)$, $I_1^q(y)=-c_V^e c_A^q A(y)+c_A^e c_V^q B(y)$, $\chi=Q^4 /[(Q^2-M_Z^2)^2+\Gamma_Z^2 M_Z^2] \sin ^4 2 \theta_W$, $\chi_{\text {int }}^q=-2 e_q Q^2(Q^2-M_Z^2) /[(Q^2-M_Z^2)^2+\Gamma_Z^2 M_Z^2] \sin ^2 2 \theta_W$, $B(y)=1-2y$, $c_1^e=(c_V^e)^2+(c_A^e)^2$, $c_3^e=2c_V^e c_A^e$, $c_1^q=(c_V^q)^2+(c_A^q)^2$ and $c_3^q=2c_V^q c_A^q$. Notice that $\omega_{\bar q} (y) = \omega_{q} (1-y)$ and $\Delta\omega_{\bar q} = - \Delta\omega_q (1-y)$.

The above result has a simple physical interpretation in the parton model. We use $d\hat \sigma_{\lambda_q \lambda_{\bar q}}/dy$ to denote the partonic cross section where $\lambda_{q/\bar q} = \pm 1$ specifies the helicity of the fragmenting quark or antiquark. There are in total four combinations, i.e., $\hat \sigma_{+-}$, $\hat \sigma_{-+}$, $\hat \sigma_{++}$, and $\hat \sigma_{--}$. To bridge the partonic cross section to the hadronic one, we need to employ the probability density of producing hadrons with helicity $\lambda_h$ from quark with helicity $\lambda_q$, which is denoted as ${\cal D}_{q} (z,\lambda_h; \lambda_{q})$. It is parametrized as ${\cal D}_{q} (z,\lambda_h; \lambda_{q}) = D_{1,q} (z) + \lambda_q \lambda_h  G_{1L,q} (z)$. Therefore, we find that $2G_{1L,q} (z) = {\cal D}_{q} (z, + ; +) -  {\cal D}_{q} (z, - ; +) = {\cal D}_{q} (z, - ; -) -  {\cal D}_{q} (z, +; -)$ is just the longitudinal spin transfer and $2D_{1,q} (z) = {\cal D}_{q} (z, + ; +) +  {\cal D}_{q} (z, - ; +)  = {\cal D}_{q} (z, + ; -) +  {\cal D}_{q} (z, - ; -)$ is the unpolarized FF. This approach is also known as the helicity amplitude approach, which has been widely used in literatures, e.g., Refs.~\cite{Anselmino:2005sh, Anselmino:2011ch, DAlesio:2021dcx, DeCausmaecker:1981wzb}. The helicity amplitudes of the most commonly used partonic processes have been computed and summarized in Ref.~\cite{Gastmans:1990xh}. The above discussion also applys to the antiquark case. 

At the end of the day, the cross section at the hadronic level is given by
\begin{align}
\frac{d\sigma}{dy dz_1 dz_2} 
& = 
  \frac{d\hat\sigma_{++}}{dy} [{\cal D}_{q} (z_{1},\lambda_{1}; +) {\cal D}_{\bar q} (z_{2},\lambda_{2}; +)]
+ \frac{d\hat\sigma_{--}}{dy} [{\cal D}_{q} (z_{1},\lambda_{1}; -) {\cal D}_{\bar q} (z_{2},\lambda_{2}; -)]
\nonumber\\
&
+ \frac{d\hat\sigma_{+-}}{dy} [{\cal D}_{q} (z_{1},\lambda_{1}; +) {\cal D}_{\bar q} (z_{2},\lambda_{2}; -)]
+ \frac{d\hat\sigma_{-+}}{dy} [{\cal D}_{q} (z_{1},\lambda_{1}; -) {\cal D}_{\bar q} (z_{2},\lambda_{2}; +)]
+ \{ q \leftrightarrow \bar q\},
\\
&
=
  D_{1,q} (z_1) D_{1, \bar q} (z_2) \left[ \frac{d\hat\sigma_{++}}{dy} + \frac{d\hat\sigma_{--}}{dy} + \frac{d\hat\sigma_{+-}}{dy} + \frac{d\hat\sigma_{-+}}{dy}\right]
\nonumber\\
&
+ \lambda_1 \lambda_2 G_{1L,q} (z_1) G_{1L,\bar q} (z_2) \left[ \frac{d\hat\sigma_{++}}{dy} + \frac{d\hat\sigma_{--}}{dy} - \frac{d\hat\sigma_{+-}}{dy} - \frac{d\hat\sigma_{-+}}{dy}\right]
\nonumber\\
&
+ \lambda_2 D_{1,q} (z_1) G_{1L, \bar q} (z_2) \left[ \frac{d\hat\sigma_{++}}{dy} - \frac{d\hat\sigma_{+-}}{dy} + \frac{d\hat\sigma_{-+}}{dy} - \frac{d\hat\sigma_{--}}{dy}\right]
\nonumber\\
&
+ \lambda_1 G_{1L,q} (z_1) D_{1, \bar q} (z_2) \left[ \frac{d\hat\sigma_{++}}{dy} - \frac{d\hat\sigma_{-+}}{dy} + \frac{d\hat\sigma_{+-}}{dy} - \frac{d\hat\sigma_{--}}{dy}\right]
+ \{ q \leftrightarrow \bar q\}.
\end{align}
Notice that the partonic cross section should also change accordingly, while exchanging quark and antiquark. Since $e^+e^-$ annihilation is an $s$-channel interaction, final state quark and antiquark are on the same fermionic line. We thus have $\hat \sigma_{++} = \hat \sigma_{--} = 0$ for massless quarks. To put it in another way, quark and antiquark have opposite helicities. Therefore, the above cross section reduces to following simple form 
\begin{align}
\frac{d\sigma}{dy dz_1 dz_2} 
& = 
\Bigl[D_{1,q} (z_1) D_{1,\bar q} (z_2) - \lambda_1 \lambda_2 G_{1L,q} (z_1) G_{1L,\bar q} (z_2) \Bigr] \left[ \frac{d\hat\sigma_{+-}}{dy} + \frac{d\hat\sigma_{-+}}{dy}\right]
\nonumber\\
& +
\Bigl[\lambda_1 G_{1L,q} (z_1) D_{1,\bar q} (z_2) - \lambda_2 D_{1,q} (z_1) G_{1L,\bar q} (z_2) \Bigr] \left[ \frac{d\hat\sigma_{+-}}{dy} - \frac{d\hat\sigma_{-+}}{dy}\right]
+ \{ q \leftrightarrow \bar q\}.
\label{eq:eecs-helicity}
\end{align}
At LO, the remaining two nonvanishing partonic cross sections are given by
\begin{align}
&
\frac{d\hat\sigma_{+-}}{dy} =\frac{\pi N_c \alpha^2}{Q^2} \sum_q 
\Bigl[ 
e_q^2 A(y) +\chi_{\text {int }}^q(c_V^q-c_A^q)[c_V^eA(y)+c_A^eB(y)]
+\chi (c_1^q-c_3^q)[c_1^eA(y)+c_3^eB(y)]
\Bigr],\label{eq:eepm}\\
&
\frac{d\hat\sigma_{-+}}{dy} =\frac{\pi N_c \alpha^2}{Q^2} \sum_q 
\Bigl[ 
e_q^2 A(y) +\chi_{\text {int }}^q(c_V^q+c_A^q)[c_V^eA(y)-c_A^eB(y)]
+\chi (c_1^q+c_3^q)[c_1^eA(y)-c_3^eB(y)]
\Bigr].\label{eq:eemp}
\end{align}
Substituting Eqs. (\ref{eq:eepm}-\ref{eq:eemp}) into \ref{eq:eecs-helicity}, it is straightforward to varify that this calculation based on the probability density interpretation exactly reproduces that from the LO and LT approximation in Eq.~(\ref{eq:cross-ee}). This is not surprising, since the physical interpretations of LT PDFs and FFs coincide with the probability densities. However, this approach cannot go beyond LT, since higher twist PDFs or FFs cannot be interpreted as probability densities. As shown in Refs.~\cite{Mulders:1995dh, Boer:1997mf, Boer:2003cm, Bacchetta:2006tn, Pitonyak:2013dsu, Wei:2014pma, Metz:2016swz}, higher twist PDFs and FFs are convoluted with different hard coefficients. 

Before presenting the numerical results, we make one more comment concerning the vanishing cross sections (i.e., $\hat \sigma_{++}$ and $\hat \sigma_{--}$). They vanish because that the two final state partons are on the same trace line. This feature maximizes the longitudinal polarization correlation. As discussed in the next section, they also vanish in several similar channels (such as $g + g\to q + \bar q$) in $pp$ collisions. However, for the other channels, such as $q_i + q_j \to q_i + q_j$, we have nonzero $\hat \sigma_{++}$ and $\hat \sigma_{--}$. Therefore, the hard coefficient of $G_{1L} (z_1)G_{1L} (z_2)$ and that of $D_1 (z_1) D_1 (z_2)$ are no longer the same.

It is straightforward to obtain the dihadron polarization correlation from Eq.~(\ref{eq:cross-ee}), which reads
\begin{align}
{\cal C}_{LL} (y,z_1,z_2) = - 
\frac{
\sum_q \omega_q (y) G_{1L,q} (z_1) G_{1L,\bar q} (z_2) + \{q \leftrightarrow \bar q\}
}{
\sum_q \omega_q (y) D_{1,q} (z_1) D_{1,\bar q} (z_2) + \{q \leftrightarrow \bar q\}
}.
\end{align}
Integrating over $y$, we obtain
\begin{align}
{\cal C}_{LL} (z_1,z_2) = - 
\frac{
\sum_q (c_1^e c_1^q \chi + e_q^2 + \chi_{\rm int}^q c_V^e c_V^q) [ G_{1L,q}^{\Lambda} (z_1) G_{1L,\bar q}^{\bar\Lambda} (z_2) + G_{1L,\bar q}^{\Lambda} (z_1) G_{1L,q}^{\bar\Lambda} (z_2) ]
}{
\sum_q (c_1^e c_1^q \chi + e_q^2 + \chi_{\rm int}^q c_V^e c_V^q) [D_{1,q}^{\Lambda} (z_1) D_{1,\bar q}^{\bar\Lambda} (z_2) + D_{1,q}^{\bar\Lambda} (z_1) D_{1,\bar q}^{\Lambda} (z_2)]
},
\end{align}
where the complete formula has been explicitly laid out to avoid possible confusion.

\begin{figure}[tb] \centering
\includegraphics[width=0.4\textwidth]{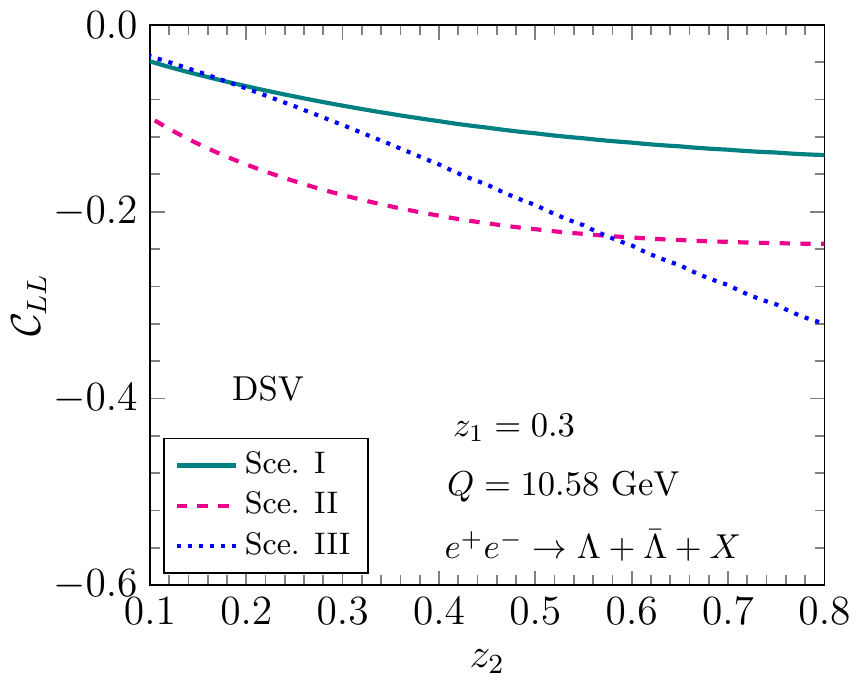}
\includegraphics[width=0.4\textwidth]{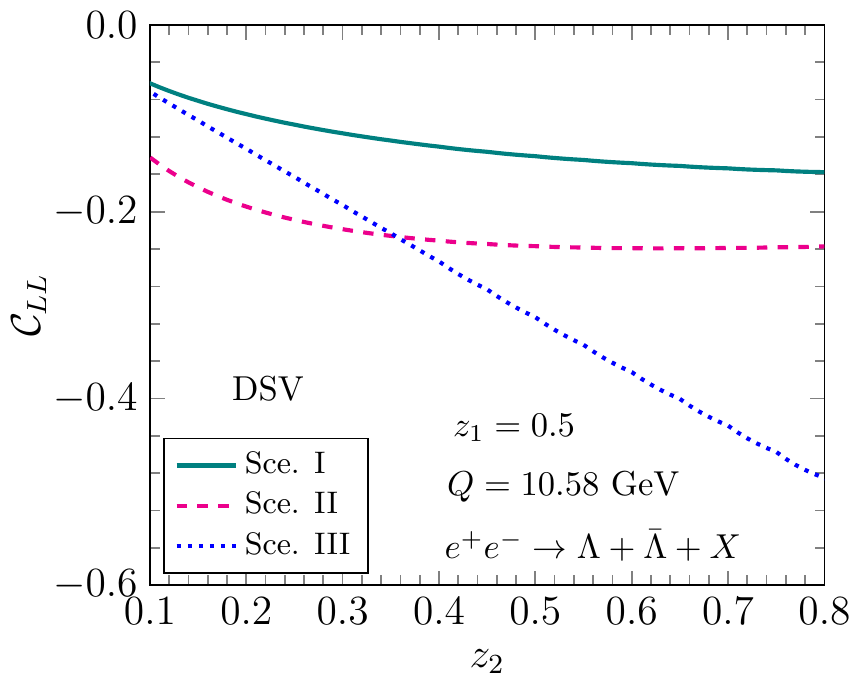}
\\
\includegraphics[width=0.4\textwidth]{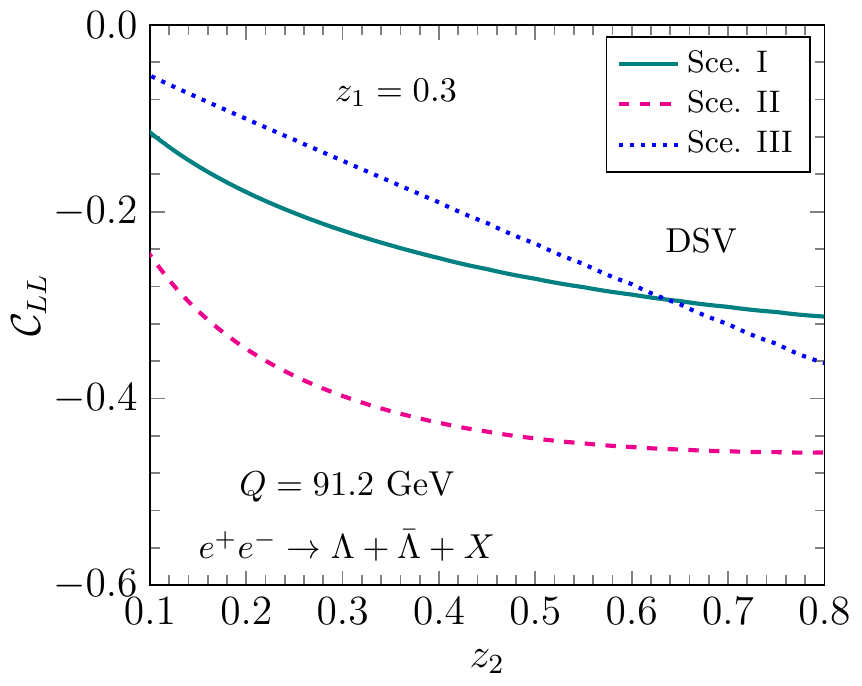}
\includegraphics[width=0.4\textwidth]{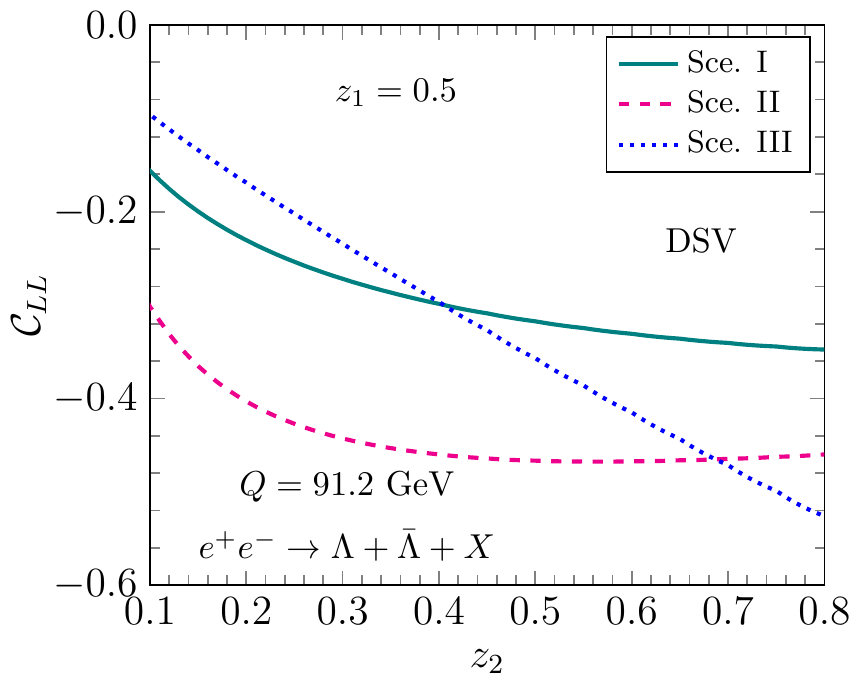}
\caption{Predictions of dihadron polarization correlation in $e^+e^-$ annihilations at Belle and LEP energies.}
\label{fig:ee}
\end{figure}

Since the $\Lambda$ FFs in the market, such as DSV \cite{deFlorian:1997zj} and AKK08 \cite{Albino:2008fy}, do not distinguish $\Lambda$ from $\bar \Lambda$ in the unpolarized FFs, we need to employ the following prescription in phenomenology
\begin{align}
&
D_{1,q}^{\Lambda} (z) = D_{1,\bar q}^{\bar \Lambda} (z) = \frac{1+z}{2} D_{1, q}^{\Lambda + \bar \Lambda} (z),
\\
&
D_{1,\bar q}^{\Lambda} (z) = D_{1,q}^{\bar \Lambda} (z) = \frac{1-z}{2} D_{1, q}^{\Lambda + \bar \Lambda} (z).
\end{align}
This approximation has also been employed in Refs.~\cite{Chen:2021hdn, Chen:2021zrr}. We have numerically tested that such a prescription can describe the longitudinal polarization of single inclusive $\Lambda$ production measured by the ALEPH \cite{ALEPH:1996oew} and OPAL \cite{ OPAL:1997oem} collaborations at LEP. 

Employing the DSV parametrization \cite{deFlorian:1997zj} for polarized and unpolarized FFs, we present our numerical predictions for the dihadron polarization correlation in $e^+e^-$ annihilations at Belle and LEP energies with several typical kinematic values in Fig.~\ref{fig:ee}. It clearly shows that the polarization correlation at the Belle energy, roughly speaking, has a similar magnitude with that at the LEP energy. Such a significant polarization correlation at different collisional energies makes it possible to extract the longitudinal spin transfer from the Belle experiment where the electromagnetic interaction dominates.

\section{Dihadron polarization correlation in $pp$ collisions}

In principle, the longitudinal spin transfer can also be probed in the polarized SIDIS \cite{E665:1999fso, HERMES:1999buc, HERMES:2006lro, COMPASS:2009nhs} and polarized $pp$ collisions \cite{STAR:2009hex, STAR:2018pps}. RHIC is the only polarized $pp$ collider so far. It has measured the longitudinal spin transfer coefficient $D_{LL}$ in Refs.~\cite{STAR:2009hex, STAR:2018pps}. Such an observable probes the combination of the longitudinal spin transfer in PDFs, $g_{1L} (x)$, and that in FFs, $G_{1L} (z)$. Besides, RHIC, Tevatron, and LHC experiments have accumulated enormous experimental data in unpolarized $pp$ collisions. It will be interesting to analyze the longitudinal spin correlation of two almost back-to-back hadrons, which is sensitive to $G_{1L}(z_1) \otimes G_{1L} (z_2)$. It is free from the contamination of the longitudinal spin transfer in PDFs. Furthermore, in light of the amount of unpolarized data that have already been collected, this analysis can shed new light on the quantitative study of longitudinal spin transfer. 

To be more specific, we consider the following two processes
\begin{align}
&
p + p \to \Lambda (\lambda_1, \eta_1, \bm{p}_{T1}) + \bar \Lambda (\lambda_2, \eta_2, \bm{p}_{T2}) + X, 
\\
&
p + p \to \Lambda (\lambda_1, \eta_1, \bm{p}_{T1}) + \Lambda (\lambda_2, \eta_2, \bm{p}_{T2}) + X, 
\end{align}
where these two final state $\Lambda$ hyperons are almost back-to-back in the transverse plane (i.e., $\bm{p}_{T1}/z_1 \sim - \bm{p}_{T2}/z_2$). $\eta_{1,2}$ are the rapidities of $\Lambda$'s. Notice that in $pp$ collisions, we usually choose the beam direction as the $z$ axis. The transverse momenta here are different from those in the TMD factorization. 

We stay in the context of the collinear factorization, since the TMD factorization does not apply in this case \cite{Collins:2007nk, Rogers:2010dm}. Following the same strategy as that in $e^+e^-$ annihilations, we can relate the hadronic cross section to the partonic ones with
\begin{align}
\frac{d\sigma}{d\eta_1 d^2 \bm{p}_{T1} d\eta_2 d^2 \bm{p}_{T2}} 
& = 
\int \frac{dz_1}{z_1^2} \frac{dz_2}{z_2^2} 
\sum_{ab\to cd}
\sum_{\lambda_c\lambda_d} 
\frac{1}{\pi} \frac{d\hat\sigma_{\lambda_c\lambda_d}^{ab\to cd}}{dt} 
x_a f_{1,a}(x_a) x_b f_{1,b}(x_b)
{\cal D}_{c} (z_{1},\lambda_{1}; \lambda_c) {\cal D}_{d} (z_{2},\lambda_{2}; \lambda_d)
\nonumber
\\
& \times
\delta^2 \left( \frac{\bm{p}_{T1}}{z_1} + \frac{\bm{p}_{T2}}{z_2} \right)
+ \{ c \leftrightarrow d \},
\label{eq:cs-pp}
\end{align}
where $f_{1,a/b} (x_{a/b})$ is the collinear PDF with $x_a = \frac{p_{T1}}{z_1 \sqrt{S}} (e^{\eta_1} + e^{\eta_2})$ and $x_b = \frac{p_{T1}}{z_1 \sqrt{S}} (e^{-\eta_1} + e^{-\eta_2})$ the parton momentum fractions and $\sqrt{S}$ the center-of-mass energy of colliding protons. Again, since we cannot determine which parton fragments to the first $\Lambda$ hyperon, the exchange between $\{c \leftrightarrow d\}$ is implicit for non-identical final state partons. Here, $\sum_{ab\to cd}$ means that we need to sum over all possible partonic scatterings. They can be classified into eight channels, namely $q_i + q_j \to q_i + q_j$, $q_i + q_i \to q_i + q_i$, $q_i + \bar q_i \to q_i + \bar q_i$, $q_i + \bar q_i \to q_j + \bar q_j$, $g + g \to q_i + \bar q_i$, $q_i + \bar q_i \to g + g$, $q + g \to q + g$, and $g + g \to g + g$ channels with $q_i$ and $q_j$ representing quarks with different flavors and $g$ denoting a gluon. At LO, it is direct to evaluate these helicity-dependent cross sections. The helicity amplitudes of all different partonic channels can be find in Ref.~\cite{Gastmans:1990xh}. After summing over the spin of initial state partons, we obtain the partonic cross sections used in our approach. We present the formula explicitly as follows.

For the $g + g \to q_i + \bar q_i$ and $q_i + \bar q_i \to q_j + \bar q_j$ channels, the final state $q\bar q$ pair is connected. Therefore, $\hat \sigma_{++}$ and $\hat \sigma_{--}$ disappear. We find \cite{Gastmans:1990xh}
\begin{align} 
&
\frac{d\hat{\sigma}^{g + g \to q_i + \bar q_i}_{+-}}{dt} = \frac{d\hat{\sigma}^{g + g \to q_i + \bar q_i}_{-+}}{dt} = \frac{\pi \alpha_{s}^{2}}{12s^{2}}\left[\frac{t}{u}+\frac{u}{t} -\frac{9}{4}\frac{t^2+u^2}{s^2}\right],
\\
& \frac{d\hat{\sigma}^{q_i + \bar q_i \to q_j +\bar q_j}_{+-}}{dt} = \frac{d\hat{\sigma}^{q_i + \bar q_i \to q_j +\bar q_j}_{-+}}{dt} = \frac{2\pi \alpha_{s}^{2}}{9 s^{2}}\left[\frac{t^{2}+u^{2}}{s^{2}}\right].
\end{align}
Here, we have used the partonic Mandelstam variables which are defined as $s = x_a x_b S$, $t = - x_a \sqrt{S} \frac{p_{T1}}{z_1} e^{-\eta_1}$ and $u=- x_a \sqrt{S} \frac{p_{T1}}{z_1} e^{-\eta_2}$. Summing over the helicity of final state partons, we arrive at the spin-summed cross sections which agree with those summarized in Ref.~\cite{Owens:1986mp}.

The $q_i + \bar q_i \to q_i + \bar q_i$, $q_i +  q_i \to q_i +  q_i$, and $q_i + q_j \to q_i + q_j (i\neq j)$ channels are more involved. All four helicity-dependent cross sections contribute. They are given by \cite{Gastmans:1990xh}
\begin{align}
& \frac{d\hat{\sigma}^{q_i + \bar q_i \to q_i + \bar q_i}_{++}}{dt} = \frac{d\hat{\sigma}^{q_i + \bar q_i \to q_i + \bar q_i}_{--}}{dt} = \frac{2\pi \alpha_{s}^{2}}{9 s^{2}} \frac{s^{2}}{t^{2}},
\\
& \frac{d\hat{\sigma}^{q_i + \bar q_i \to q_i + \bar q_i}_{+-}}{dt} = \frac{d\hat{\sigma}^{q_i + \bar q_i \to q_i + \bar q_i}_{-+}}{dt} = \frac{2\pi \alpha_{s}^{2}}{9 s^{2}}\left[\frac{u^{2}}{t^{2}}+\frac{t^{2}+u^{2}}{s^{2}}-\frac{2}{3}\frac{u^{2}}{st}\right],
\\
& \frac{d\hat{\sigma}^{q_i + q_i \to q_i + q_i}_{++}}{dt} = \frac{d\hat{\sigma}^{q_i + q_i \to q_i + q_i}_{--}}{dt} = \frac{2\pi \alpha_{s}^{2}}{9 s^{2}}\left[\frac{s^{2}}{u^{2}}+\frac{s^{2}}{t^{2}}-\frac{2}{3}\frac{s^{2}}{tu}\right],
\\
& \frac{d\hat{\sigma}^{q_i + q_i \to q_i + q_i}_{+-}}{dt} = \frac{d\hat{\sigma}^{q_i + q_i \to q_i + q_i}_{-+}}{dt} = \frac{2\pi \alpha_{s}^{2}}{9 s^{2}}\left[\frac{t^{2}}{u^{2}}+\frac{u^{2}}{t^{2}}\right],
\\
& \frac{d\hat{\sigma}^{q_i + q_j \to q_i + q_j}_{++}}{dt} = \frac{d\hat{\sigma}^{q_i + q_j \to q_i + q_j}_{--}}{dt} = \frac{2\pi \alpha_{s}^{2}}{9 s^{2}}\frac{s^{2}}{t^{2}},
\\
& \frac{d\hat{\sigma}^{q_i + q_j \to q_i + q_j}_{+-}}{dt} = \frac{d\hat{\sigma}^{q_i + q_j \to q_i + q_j}_{-+}}{dt} = \frac{2\pi \alpha_{s}^{2}}{9 s^{2}}\frac{u^{2}}{t^{2}}.
\end{align}
Adding $\hat \sigma_{++}$, $\hat\sigma_{--}$, $\hat\sigma_{+-}$, and $\hat\sigma_{-+}$ together, we arrive at the spin-summed partonic cross section for each channel, which agrees with that in Ref.~\cite{Owens:1986mp}. Furthermore, we always have $\hat \sigma_{++} + \hat \sigma_{+-} - \hat \sigma_{-+} - \hat \sigma_{--} = 0$ and $\hat \sigma_{++} + \hat \sigma_{-+} - \hat \sigma_{+-} - \hat \sigma_{--} = 0$ for all the channels. This is expected since they correspond to the helicity of parton $c$ and $d$ respectively while averaging over the helicity of the other final state parton. Single inclusive partons are not longitudinally polarized in unpolarized $pp$ collisions. Last but not least, all those helicity-dependent cross sections are positive definite. Therefore, the polarization correlations of these three channels are expected to be smaller than those of $g + g \to q_i + \bar q_i$ and $q_i + \bar q_i \to q_j + \bar q_j$ channels.

The other channels involve one or two final state gluons. The longitudinal spin transfer of $g\to \Lambda$ is assumed to be negligiable at the initial factorization scale in the DSV parameterization \cite{deFlorian:1997zj}, since it is poorly constrained by the $e^+e^-$ data. Nonetheless, it can still accumulate contributions through DGLAP evolution and becomes important at higher factorization scale. In contrast with the $e^+e^-$ case, the gluon channel in $pp$ collisions plays a crucial role in the $\Lambda$ hyperon production, particularly at the LHC energy. Therefore, the experimental data from $pp$ collisions is vital on removing ambiguities in the gluon spin transfer. The partonic cross sections with circularly polarized gluons are given by \cite{Gastmans:1990xh}
\begin{align}
& \frac{d\hat{\sigma}^{g + q_i \to g + q_i}_{++}}{dt} = \frac{d\hat{\sigma}^{g + q_i \to g + q_i}_{--}}{dt} = \frac{\pi \alpha_{s}^{2}}{2 s^{2}}\left[\frac{s^{2}}{t^{2}}-\frac{4s}{9u}\right],
\\
& \frac{d\hat{\sigma}^{g + q_i \to g + q_i}_{+-}}{dt} = \frac{d\hat{\sigma}^{g + q_i \to g + q_i}_{-+}}{dt} = \frac{\pi \alpha_{s}^{2}}{2 s^{2}}\left[\frac{u^{2}}{t^{2}}-\frac{4u}{9s}\right],
\\
& \frac{d\hat{\sigma}^{q_i + \bar{q_i} \to g + g}_{++}}{dt} = \frac{d\hat{\sigma}^{q_i + \bar{q_i} \to g + g}_{--}}{dt} = 0,
\\
& \frac{d\hat{\sigma}^{q_i + \bar{q_i} \to g + g}_{+-}}{dt} = \frac{d\hat{\sigma}^{q_i + \bar{q_i} \to g + g}_{-+}}{dt} = \frac{4\pi \alpha_{s}^{2}}{3 s^{2}}\left[\frac{4u}{9t}+\frac{4t}{9u}-\frac{t^{2}+u^2}{s^2}\right],
\\
& \frac{d\hat{\sigma}^{g+g\to g+g}_{++}}{dt} = \frac{d\hat{\sigma}^{g+g\to g+g}_{--}}{dt} = \frac{9\pi \alpha_{s}^{2}}{16 s^{2}}\frac{s^2}{t^2u^2}(s^2+t^2+u^2),
\\
& \frac{d\hat{\sigma}^{g+g\to g+g}_{+-}}{dt} = \frac{d\hat{\sigma}^{g+g\to g+g}_{-+}}{dt} = \frac{9\pi \alpha_{s}^{2}}{16 s^{2}} \left[\frac{u^2}{s^2 t^2}+\frac{t^2}{s^2u^2}\right](s^2+t^2+u^2).
\end{align}

Substituting the decomposition of ${\cal D}_{c,d}$ into Eq.~(\ref{eq:cs-pp}), we arrive at the cross section in terms of the unpolarized FF and the longitudinal spin transfer. It reads
\begin{align}
\frac{d\sigma}{d\eta_1 d\eta_2} = & \int dz_1dz_2 \frac{1}{z_1^2} \int d^2 \bm{p}_{T1} \sum_{ab\to cd} 
x_a f_{1,a}(x_a) x_b f_{1,b}(x_b) \frac{1}{\pi}
\nonumber\\
&
\times 
\Biggl\{
D_{1,c} (z_1) D_{1,d} (z_2) \Bigl[ \frac{d\hat\sigma_{++}^{ab\to cd}}{dt} + \frac{d\hat\sigma_{--}^{ab\to cd}}{dt} + \frac{d\hat\sigma_{+-}^{ab\to cd}}{dt} + \frac{d\hat\sigma_{-+}^{ab\to cd}}{dt}  \Bigr]
\nonumber\\
&
+ \lambda_1 \lambda_2 G_{1L,c} (z_1) G_{1L,d} (z_2) 
\Bigl[ \frac{d\hat\sigma_{++}^{ab\to cd}}{dt} + \frac{d\hat\sigma_{--}^{ab\to cd}}{dt} - \frac{d\hat\sigma_{+-}^{ab\to cd}}{dt} - \frac{d\hat\sigma_{-+}^{ab\to cd}}{dt}  \Bigr]
\Biggr\}
+ \{c \leftrightarrow d\}.
\end{align}
Thus, the dihadron polarization correlation in $pp$ collisions is given by 
\begin{align}
{\cal C}_{LL} (\eta_1, \eta_2) = \frac{
\int dz_1 dz_2 \int \frac{d^2 \bm{p}_{T1}}{z_1^2} \sum_{ab\to cd} x_a f_{1,a}(x_a) x_b f_{1,b}(x_b) \frac{1}{\pi} \frac{d\hat\sigma^{\rm dif}}{dt} G_{1L,c} (z_1) G_{1L,d} (z_2) + \{c \leftrightarrow d\}
}{
\int dz_1 dz_2 \int \frac{d^2 \bm{p}_{T1}}{z_1^2} \sum_{ab\to cd} x_a f_{1,a}(x_a) x_b f_{1,b}(x_b) \frac{1}{\pi} \frac{d\hat\sigma^{\rm sum}}{dt} D_{1,c} (z_1) D_{1,d} (z_2) + \{c \leftrightarrow d\}
} ,
\end{align}
where we have used the following shorthand notation,
\begin{align}
& \frac{d\hat\sigma^{\rm dif}}{dt} = \frac{d\hat\sigma_{++}^{ab\to cd}}{dt} + \frac{d\hat\sigma_{--}^{ab\to cd}}{dt} - \frac{d\hat\sigma_{+-}^{ab\to cd}}{dt} - \frac{d\hat\sigma_{-+}^{ab\to cd}}{dt},
\\
& \frac{d\hat\sigma^{\rm sum}}{dt} = \frac{d\hat\sigma_{++}^{ab\to cd}}{dt} + \frac{d\hat\sigma_{--}^{ab\to cd}}{dt} + \frac{d\hat\sigma_{+-}^{ab\to cd}}{dt} + \frac{d\hat\sigma_{-+}^{ab\to cd}}{dt}.
\end{align}

\begin{figure}[tb]
\includegraphics[width=0.4\textwidth]{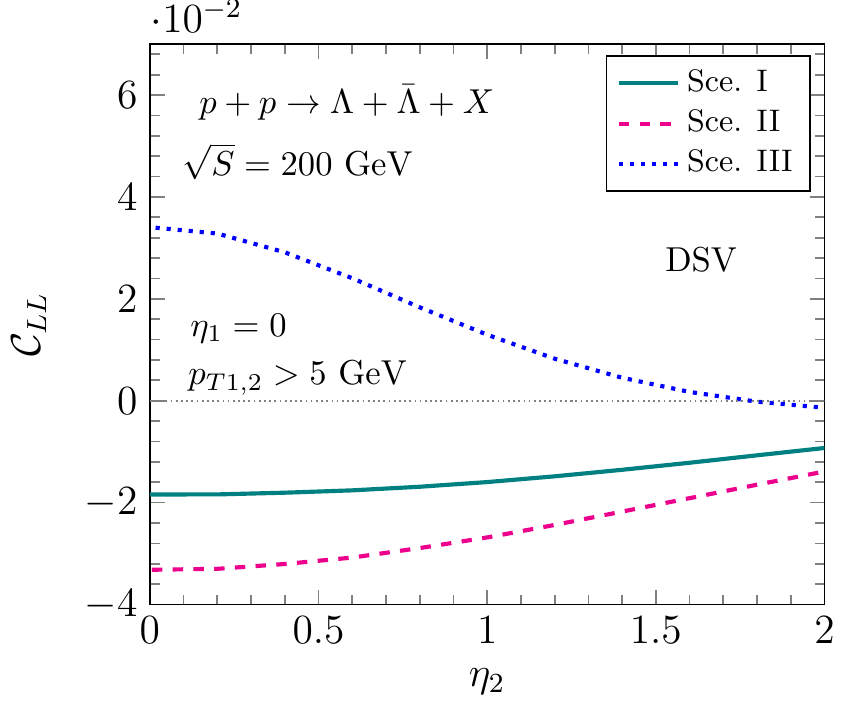}
\includegraphics[width=0.4\textwidth]{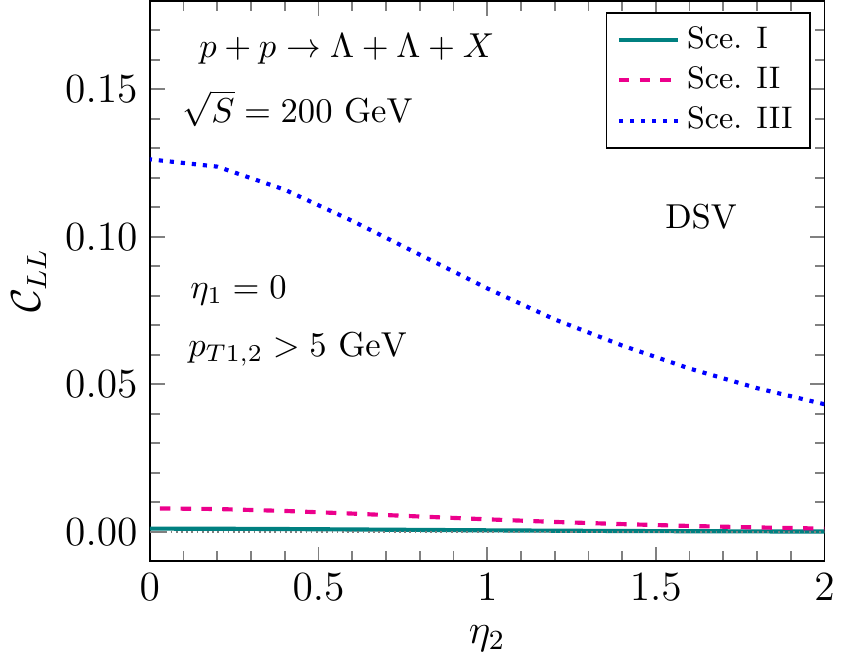}
\includegraphics[width=0.4\textwidth]{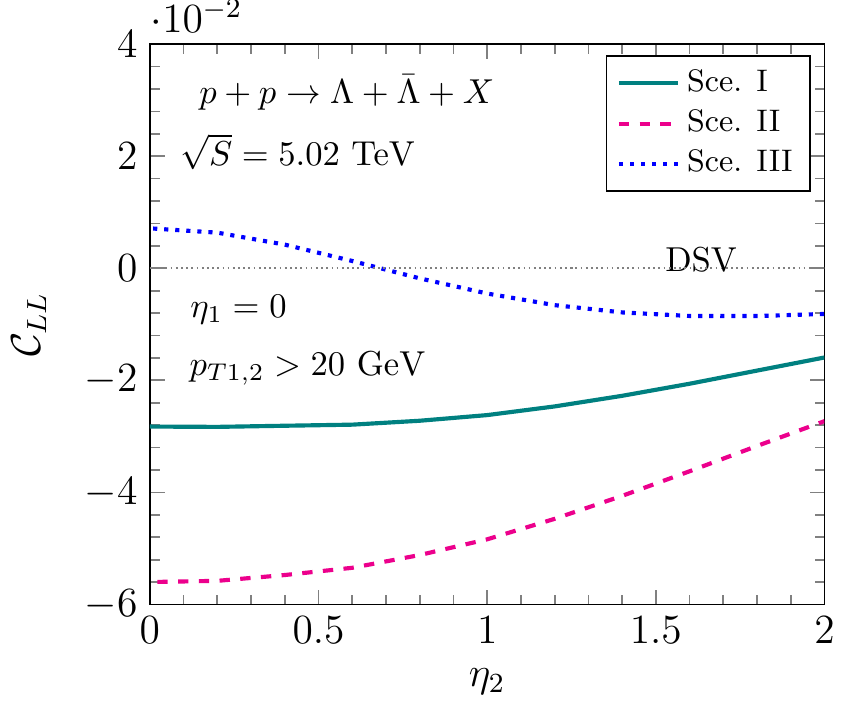}
\includegraphics[width=0.4\textwidth]{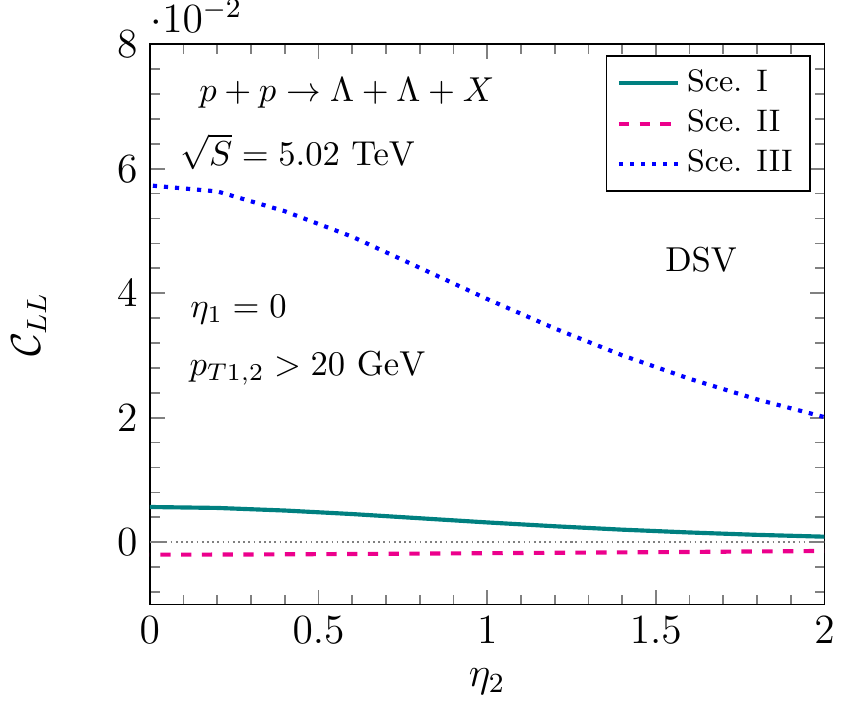}
\includegraphics[width=0.4\textwidth]{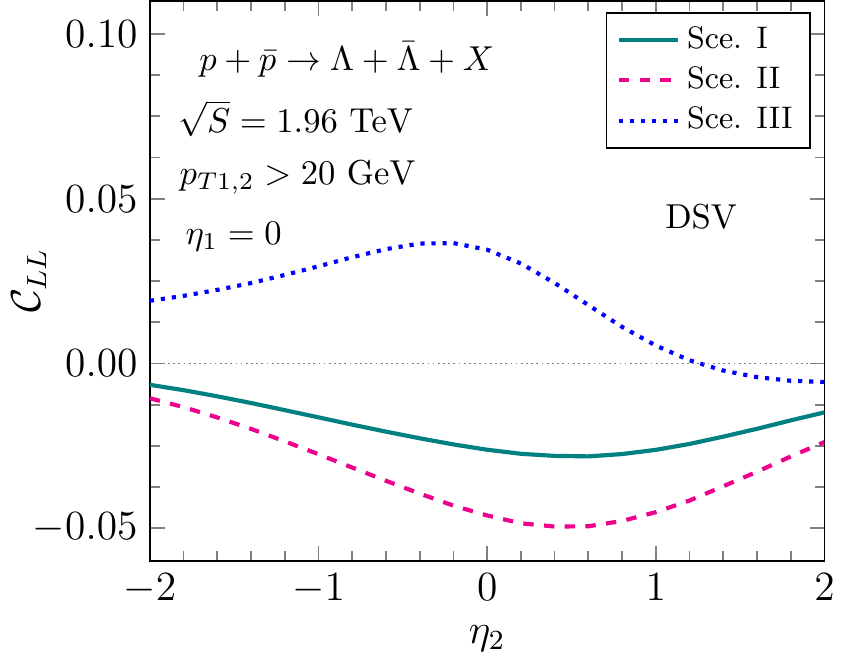}
\includegraphics[width=0.4\textwidth]{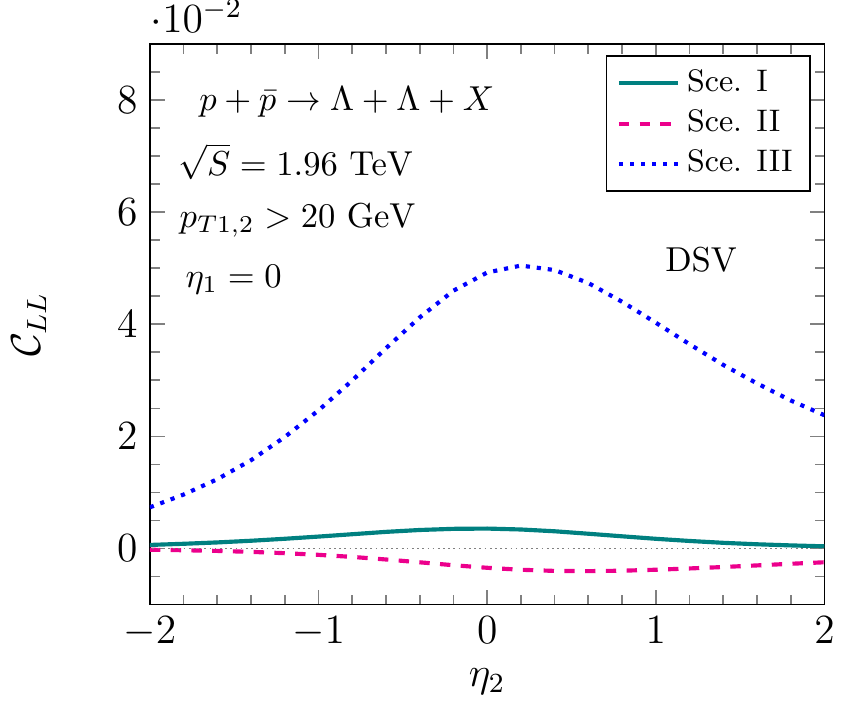}
\caption{Predictions for the dihadron polarization correlation in $pp/p\bar p$ collisions at RHIC, LHC, and Tevatron energies.}
\label{fig:pp}
\end{figure}

We use $\eta_1 \sim \eta_2 \sim 0$ as an example to demonstrate some interesting features. In this approximation,  $t \sim u \sim - s/2$. We can then perform a quick evaluation and figure out the sign of the polarization correlation at the partonic level. Notice that this approximation is only employed to facilitate a quick estimate. The conclusion does not change as long as $s+t+u=0$ and $t,u < 0$. Assuming that the longitudinal spin transfers are always positive for all partons, we then find that the $q_i + \bar q_i \to q_j + \bar q_j$, $q_i + \bar q_i \to g + g$, and $g+g\to q_i + \bar q_i$ channels contribute to negative dihadron polarization correlations and the other channels offer positive contributions. Loosely speaking, $s$-channels lead to a negative correlation, while $t$-channels lead to a positive correlation. Therefore, there is a partial cancellation among different partonic channels, which results in a much smaller dihadron polarization correlation in $pp$ collisions vis-a-vis that in $e^+e^-$ annihilations.  Notice that this assumption is not always the case. The longitudinal spin transfers of $u$ and $d$ quarks are assumed to be small but negative in the second scenario in the DSV parameterizations \cite{deFlorian:1997zj}. 

Utilizing CTEQ PDFs \cite{arXiv:1506.07443} and DSV FFs \cite{deFlorian:1997zj}, we compute the dihadron polarization correlation in $pp$ and $p\bar p$ collisions at different collisional energies and show the numerical results in Fig.~\ref{fig:pp}. The cross section in the middle rapidity is much larger than that in the forward rapidity. For the sake of more statistics, we present our predictions for $\eta_1=0$ and $|\eta_2|<2$. Our predictions for $pp$ collisions are symmetric between positive and negative $\eta_2$. Therefore, we only show the positive half in our predictions for RHIC and LHC experiments. For those typical kinematic configurations, we find that the dihadron polarization correlation is about a few percent. It is much smaller than that in $e^+e^-$ annihilations. The reason is twofold. First, the final state quarks in $e^+e^-$ annihilations are always connected, which maximizes the polarization correlation. In contrast, in $pp$ collisions, besides the $s$-channels which lead to the negative polarization correlation, there are also sizable contributions from the $t$-channels which give birth to the positive polarization correlation. As mentioned above, the partial cancellation reduces the polarization correlation at the partonic level, which translates into smaller correlation at the hadronic level. Second, the spin transfer of a circularly polarized gluon is assumed to be negligible at the initial scale in the DSV parameterization \cite{deFlorian:1997zj}. The contribution only arises from the DGLAP evolution. Therefore, the numerical calculation underestimates the polarization correlation. This work can be viewed as the lower bound. A rigorous investigation on the circularly polarized gluon FF (a.k.a. the gluon spin transfer) can be carried out once we had more experimental data. 

By increasing the collisional energy, the momentum fractions, $x_1$ and $x_2$, decrease. In general, we expect more gluon contributions. To be comparable with each other, we elevate the threshold for the transverse momentum of $\Lambda$ from 5 GeV to 20 GeV while moving from the RHIC energy to Tevatron or LHC energies. On the bright side, we have more phase space to investigate the $p_T$ dependence at higher collisional energy. The future measurements of the polarization correlation at different $p_T$ ranges can be employed to probe the difference between quark and gluon spin transfers.

Furthermore, in Scenario I of the DSV parametrization \cite{deFlorian:1997zj}, the spin transfer comes almost exclusively from $s \to \Lambda$. Therefore, the dihadron polarization correlation in the $p+p\to \Lambda + \bar\Lambda +X$ process is mainly determined by the polarization correlation of the $s\bar s$ pair. Since the production of the $s \bar s$ pair is an $s$-channel dominant process, we always have negative correlations in Scenario I at different collisional energies as shown in Fig.~\ref{fig:pp}. However, the dihadron polarization correlation of the $p+p\to \Lambda +\Lambda+X$ process is mainly determined by that of the $ss$ pair, which can only be produced through the $t$-channel. Thus, the dihadron polarization correlation is then positive for this case. In Scenario III of the DSV parametrization \cite{deFlorian:1997zj}, $u$, $d$, and $s$ quarks contribute equally to the longitudinal spin transfer. For the dihadron polarization correlation of $\Lambda+\Lambda$ production, $t$-channels still dominate. Therefore, the polarization correlation is always positive. However, for that of $\Lambda+\bar\Lambda$ production, the dominant contribution varies with the kinematics. Moreover, the discussion for the second scenario is more complicated, since the longitudinal spin transfer is no longer positive for $u$ and $d$ quarks. The polarization correlation is no longer positive definite even in $\Lambda + \Lambda$ productions where $t$-channel scatterings always dominate.

\section{Summary}

In this work, we explore the opportunity of studying the longitudinal spin transfer, $G_{1L}$, through dihadron polarization correlations in $e^+e^-$ annihilations and unpolarized $pp$ collisions. Working with the LO and LT approximation, this polarization correlation can be related to the partonic helicity amplitudes. Thus, the physical picture of this observable is pretty clear. 

Employing the DSV parametrization \cite{deFlorian:1997zj} for the polarized and unpolarized FFs of $\Lambda$, we estimate that the dihadron polarization correlation is about $20\% \sim 40\%$ in $e^+e^-$ annihilations and about a few percent in $pp$ collisions. There are two reasons for the much smaller correlation in $pp$ collisions. The first one is that $s$-channels and $t$-channels contribute to the dihadron polarization correlation with different signs. The partial cancellation gives birth to a weaker polarization correlation in $pp$ collisions than that in $e^+e^-$ annihilations. The second one lies in the numerical evaluation. In all three scenarios of the DSV parametrization \cite{deFlorian:1997zj}, the longitudinal gluon spin transfer is always assumed to be negligible at the initial scale. Therefore, the numerical calculation underestimates the dihadron polarization correlation in $pp$ collisions. The ambiguity in the gluon spin transfer can only be removed by the experimental measurements. We expect that measuring dihadron polarization correlation in $pp$ collisions can play an important role on this front. It will cast new light on the fragmentation of circularly polarized gluons.

Measuring the dihadron polarization correlation in the experiment does not require polarized beams or colliding at the $Z^0$-pole. Therefore, it can be easily performed at Belle, RHIC, Tevatron, and the LHC. As a matter of fact, since most of the experimental data can be used in this analysis, we expect small statistical errors. We believe this experiment can significantly boost the quantitative study of the polarized FFs. Moreover, combining the experimental data from $e^+e^-$ and $pp$ experiments, we can better constrain the gluon spin transfer, $G_{1L, g}$. Thus, it can improve our understanding on the hadronization mechanism.

\begin{acknowledgments}

We thank K.B. Chen, T.B. Liu, and Y.K. Song for fruitful discussions. S.Y. Wei is supported by the Taishan fellowship of Shandong Province for junior scientists.

\end{acknowledgments}

\appendix

\section{Measuring dihadron polarization correlation in the experiments}

In this appendix, we show how the dihadron polarization correlation can be measured in the experiment. We consider the almost back-to-back $\Lambda$ and $\bar \Lambda$ pair production. We use $\theta_{1,2}^*$ to denote the angle between the momentum of daughter $p/\bar p$ and that of the parent $\Lambda/\bar\Lambda$ in the rest frame of $\Lambda/\bar\Lambda$ and use ${\cal P}_{++,--,+-,-+}$ to denote the probability of the four combinations of polarization states of $\Lambda$ and $\bar \Lambda$. The nomalization is given by 
\begin{align}
{\cal P}_{++} + {\cal P}_{--} + {\cal P}_{+-} + {\cal P}_{-+} = 1.
\end{align}
Thus, the dihadron polarization correlation is defined as
\begin{align}
{\cal C}_{LL} = ({\cal P}_{++} + {\cal P}_{--}) - ({\cal P}_{+-} + {\cal P}_{-+}).
\end{align}
Furthermore, the polarization of $\Lambda$ hyperons is given by $P_L^{\Lambda} = {\cal P}_{++} + {\cal P}_{+-} - {\cal P}_{-+} - {\cal P}_{--}$, while that of $\bar\Lambda$ hyperons is given by $P_L^{\bar\Lambda}  = {\cal P}_{++} + {\cal P}_{-+} - {\cal P}_{+-} - {\cal P}_{--}$.

The normalized angle distribution of the daughter $p$ and $\bar p$ is 
\begin{align}
\frac{1}{N}\frac{dN}{d\cos\theta_1^* d\cos\theta_2^*} 
&
= {\cal P}_{++} \frac{1 + \alpha \cos\theta_1^*}{2} \frac{1 + \alpha \cos\theta_2^*}{2}
+ {\cal P}_{--} \frac{1 - \alpha \cos\theta_1^*}{2} \frac{1 - \alpha \cos\theta_2^*}{2}
\nonumber\\
&
+ {\cal P}_{+-} \frac{1 + \alpha \cos\theta_1^*}{2} \frac{1 - \alpha \cos\theta_2^*}{2} 
+ {\cal P}_{-+} \frac{1 - \alpha \cos\theta_1^*}{2} \frac{1 + \alpha \cos\theta_2^*}{2}
\nonumber \\
&
= \frac{1}{4} + P_L^{\Lambda} \frac{1}{4} \alpha \cos \theta_1^* 
+ P_L^{\bar\Lambda} \frac{1}{4} \alpha \cos \theta_2^* 
+ {\cal C}_{LL} \frac{1}{4} \alpha^2 \cos\theta_1^* \cos\theta_2^*,
\end{align}
where $\alpha$ is the decay parameter of $\Lambda$ hyperons. Thus, the dihadron polarization correlation can be measured by extracting the $\langle \cos\theta_1^* \cos \theta_2^* \rangle$ module.

\end{document}